\documentclass[manuscript,screen]{acmart}

\AtBeginDocument{
  }

\acmJournal{TOSEM}

\usepackage[framemethod=TikZ]{mdframed}
\usepackage{tikz}
\usepackage{xcolor}
\usepackage{color,colortbl}
\usepackage{tcolorbox}
\usepackage{xspace}
\usepackage{algorithm}
\usepackage{algorithmic}
\usepackage{multirow}
\usepackage{enumerate}
\usepackage{makecell}
\usepackage{arydshln}
\usepackage{pifont}
\usepackage{booktabs} 
\usepackage{mathtools}

\definecolor{blueish}{RGB}{250, 250, 255}
\definecolor{greenish}{RGB}{200, 255, 200}
\definecolor{redish}{RGB}{255, 200, 200}
\definecolor{highlight}{RGB}{175, 255, 100}
\definecolor{darkred}{RGB}{139, 0, 0}
\definecolor{gray95}{gray}{0.05}
\definecolor{rowgray}{RGB}{224, 224, 224}

\newmdenv[
    tikzsetting= {fill=blueish},
    skipabove=0.33em,
    skipbelow=0.33em,
    linewidth=1pt,
    innerleftmargin=4pt,
    innerrightmargin=4pt,
    innertopmargin=2pt,
    innerbottommargin=2pt,
    linecolor=gray95,
    roundcorner=2pt, 
    shadowsize=4pt,
    shadowcolor=gray95
]{answerbox}

\newenvironment{result}
{\begin{answerbox}}
{\end{answerbox}}

\newcommand{\RS}[2]{
    \begin{result}
        \textbf{Summary of RQ#1:~}{ #2}%
    \end{result}
}

\newcommand{\ourapproach}{{\textsc DEED}\xspace}
\newcommand{\ourapproachbf}{{\textsc \textbf{DEED}}\xspace}
\newcommand{\selfrefine}{{\textsc Self-Revise}\xspace}
\newcommand{\selfrefinebf}{{\textsc \textbf{Self-Revise}}\xspace}

\acmISBN{978-1-4503-XXXX-X/18/06}

\begin{document}

\title{Exploring Data-Efficient Adaptation of Large Language Models for Code Generation}

\author{Xue Jiang}
\email{jiangxue@stu.pku.edu.cn}
\author{Yihong Dong}
\email{dongyh@stu.pku.edu.cn}
\author{Zhiyuan Fan}
\email{zyfan043@gmail.com}
\author{Zhi Jin}
\email{zhijin@pku.edu.cn}
\author{Wenpin Jiao}
\email{jwp@sei.pku.edu.cn}
\author{Ge Li}
\email{lige@pku.edu.cn}
\authornote{Corresponding author}

\affiliation{
\institution{Key Laboratory of High Confidence Software Technologies (Peking University), Ministry of Education; School of Computer Science, Peking University, Beijing}
\country{China}
}

\begin{abstract}
Although Large Language Models (LLMs) have made significant progress in code generation, they still struggle with code generation tasks in specific scenarios. These scenarios usually necessitate the adaptation of LLMs to fulfill specific needs, but the limited training data available in practice leads to poor code generation performance. Therefore, how to effectively adapt LLMs to new scenarios with few training data is a major challenge for current code generation. In this paper, we propose a novel adaptation approach named \ourapproach, which stands for \textbf{D}ata-\textbf{E}fficient adaptation with \textbf{E}rror-\textbf{D}riven learning for code generation. \ourapproach leverages the errors made by LLMs as learning opportunities, using error revision to overcome their own shortcomings, thus achieving efficient learning. Specifically, \ourapproach involves identifying error code generated by LLMs, employing \selfrefine for code revision, optimizing the model with revised code, and iteratively adapting the process for continuous improvement. Experimental results show that, compared to other mainstream fine-tuning approaches, \ourapproach achieves superior performance with few training data, showing an average relative improvement of  46.2\% in Pass@1 on multiple code generation benchmarks. We also validate the effectiveness of \selfrefine, which generates revised code that optimizes the model more efficiently compared to the code samples from datasets. Moreover, \ourapproach consistently demonstrates strong performance across various LLMs, underscoring its applicability. 

\end{abstract}



\maketitle

\section{Introduction}
Code generation is an important technology that can improve the efficiency and quality of software development. Given the human requirement expressed in natural language, code generation allows machines to generate executable programs that satisfy this requirement.
Code generation has been a research hot topic in the fields of artificial intelligence, software engineering, and natural language processing. Recently, code generation technologies have made significant advancements in both academia and industry \cite{codex, industry, CodeLlama,Agent4code,CodeP}. In particular, large language models (LLMs) demonstrate great potential in code generation tasks \cite{codegeex,codegen,incoder,codet,CR,self-planning}. However, LLMs still face significant challenges in code generation for adapting to specific scenarios or domains \cite{ahmed2024studying, DBLP:journals/corr/abs-2312-01639}.

\begin{figure*}[t!]
\centering
\includegraphics[width=0.6\textwidth]{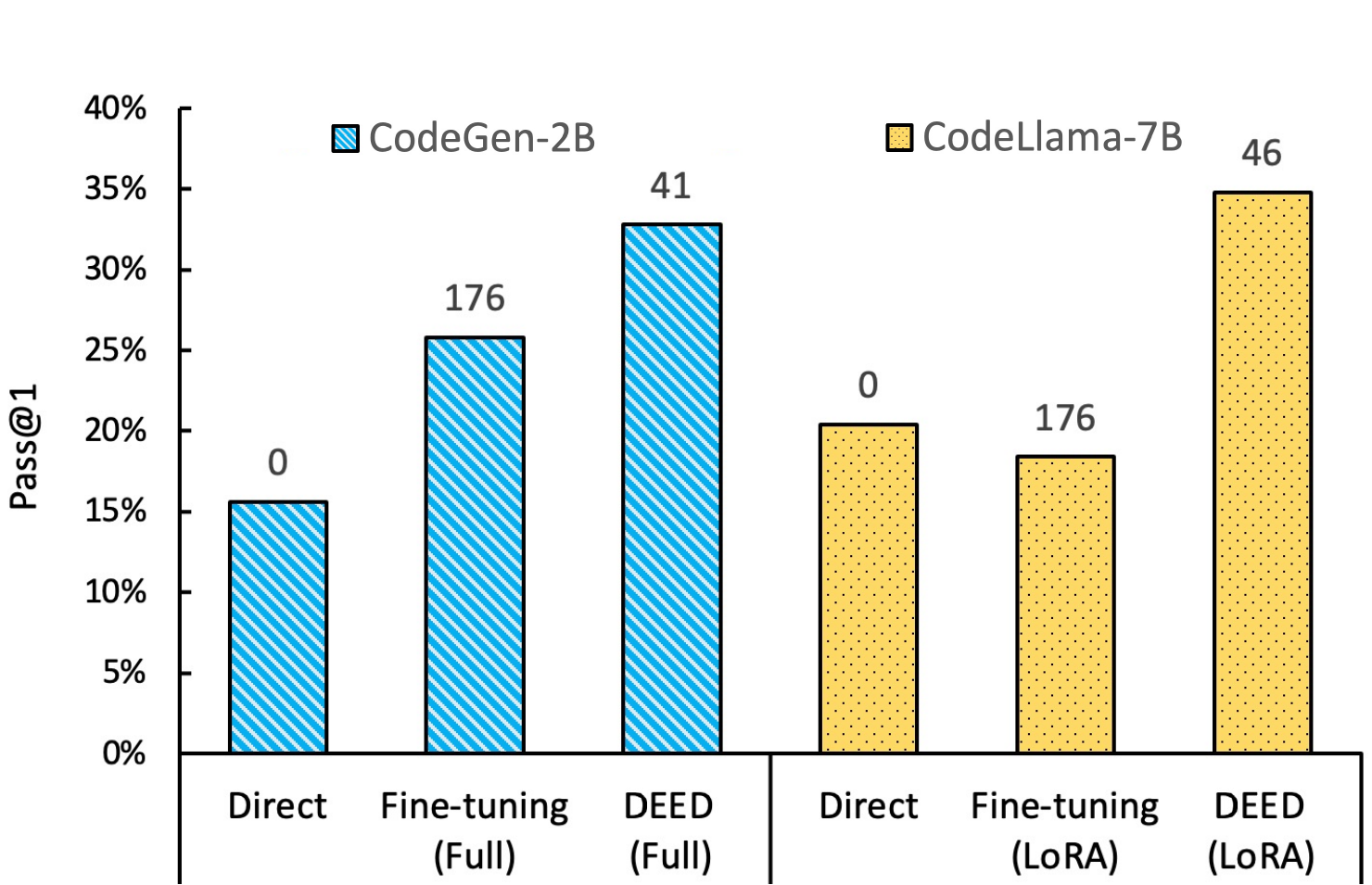}
\caption{The performance of direct generation, fine-tuning, and our proposed \ourapproach on MBPP dataset under the circumstance of limited data. The numbers on the bars indicate the training data used by different methods.}
\label{intro_fig}
\end{figure*}

For specific code generation scenarios, fine-tuning is an essential adaptation method to ensure LLMs fulfill particular needs \cite{finetune_adaptation_1,finetune_adaptation_2,finetune_adaptation_3,finetune_adaptation_4}. 
However, in these specific scenarios, it is difficult to obtain sufficient training data for fine-tuning LLMs due to common reasons such as industry secrecy and scarcity of resources. 
For example, the programs in aerospace, medical devices, and transportation, where training samples are inherently scarce due to the scarcity of high-value industry data, high collection costs, and the typical targeting of long-tail demands.
Under the circumstance of limited data, mainstream fine-tuning methods might not enable LLMs to achieve the desired code generation performance and may even lead to a degradation in model performance \cite{performance_degradation_1,performance_degradation_2, Zhang0LZZJ22}, as shown in Figure \ref{intro_fig}.
\textit{In such data-scarce scenarios, only a small amount of data is available, but high code generation performance is still required, which is hard to solve but important in real-world applications.} Consequently, how to effectively adapt LLMs to specific scenarios with limited data available is a major challenge for code generation in practice.

The mainstream fine-tuning methods use a large number of samples gathered under specific scenarios for training \cite{Gradual_Fine-Tuning}. They enable the model to exhaustively learn the features present in these samples and thus adapt to the specific scenarios.
However, they have two disadvantages. First, compelling LLMs to relearn the entire code samples of new scenarios is inefficient. Considering that LLMs are pre-trained on large-scale and diverse samples, it's reasonably assumed that they possess a certain level of general knowledge, lacking only particular information for application in specific scenarios. 
Second, when faced with insufficient data volume or data drift, the model may learn certain undesirable features (such as inaccurate or irrelevant programming knowledge and patterns), thereby affecting its learning efficiency and negatively impacting its final performance.

To overcome the disadvantages of mainstream fine-tuning methods, we take inspiration from the error-driven learning observed in humans. 1) Error-driven learning requires learners to identify their errors through testing. It helps learners to identify what they have mastered and what they still need to learn, allowing them to narrow the scope of learning and avoid wasting efforts on irrelevancies.
2) Through error revision, learners can understand their deficiencies and make targeted improvements, thus enhancing learning efficiency and effectiveness.
This motivates us to explore methods to achieve data-efficient adaptation of LLMs for code generation guided by error-driven learning.

In this paper, we propose \ourapproach, a \textbf{D}ata-\textbf{E}fficient adaptation based on \textbf{E}rror-\textbf{D}riven learning for code generation. 
\ourapproach aims to alleviate the problem of poor code generation performance of fine-tuning LLMs in data-scarce scenarios. 
Following the error-driven learning, our method proceeds in four steps:
\ding{182} \textbf{Error Code Collection.} We identify and collect error codes generated by LLMs, aiming to mine the weaknesses of LLMs. 
\ding{183} \textbf{Automatic Code Revision.} To obtain revisions of error codes in a low-cost way, we design \selfrefine to realize automatic revision leveraging information in the original dataset and code execution feedback. 
\ding{184} \textbf{Model Optimization.} 
We optimize the LLMs using the revised code, making them focus on learning the revision of these critical errors, thereby improving the learning efficiency of LLMs.
\ding{185} \textbf{Iterative Adaptation.} We adopt an iterative strategy, which involves repeating the preceding three steps, to continuously optimize and improve the performance of LLMs.

We extensively evaluate our proposed \ourapproach on five public code generation benchmarks using the test pass rate metric. Our results show that under the circumstance of limited data, \ourapproach achieves significantly better performance across various LLMs compared to the mainstream adaptation approaches. Figure \ref{intro_fig} illustrates part of this result. 
Specifically:
1) On five benchmarks, \ourapproach achieves more than 27.1\% relative improvement in the Pass@1 metric compared to the best-performing adaptation approaches. 
2) The effectiveness of \selfrefine is validated. It produces revised code that better supports the optimization of the code generation model than the code samples provided in training dataset.
3) Our iterative adaptation process facilitates continuous enhancements in performance and maintains training stability.
4) \ourapproach proves to be effective across various LLMs, demonstrating its broad applicability.
These results highlight the potential of \ourapproach in addressing the challenges of code generation in data-scarce scenarios.

To summarize, the main contributions of this paper are:
\begin{itemize}
    \item We propose that error-driven learning for LLMs adaptation is effective, i.e., utilizing revisions of LLMs' erroneous outputs for training has higher learning efficiency than samples from the dataset. 
    \item Based on the principle of error-driven learning, we propose a data-efficient adaptation method of LLMs for code generation, named \ourapproach, which can effectively adapt the model to specific scenarios with few training data.  
    \item \ourapproach outperforms the mainstream fine-tuning and prompting methods on five code generation datasets across various LLMs.
\end{itemize}
\section{Preliminary Study}
Aligning LLMs with specific scenarios and addressing their unique challenges by learning samples in the dataset is difficult when training data are limited. 
We conduct a preliminary study to explore the effectiveness of using error-driven learning in the LLMs adaptation process of code generation.

We consider the potential significant discrepancy between the model-generated output and the sample in the dataset. By learning the revisions of the model's erroneous outputs, we can find more effective navigation in the optimization process. This might provide a shorter, smoother path to a good local minimum compared to learning from samples in the dataset, rather than attempting to direct it toward a distant area that may not align well with its existing knowledge or biases.
We conduct a statistical analysis of the discrepancies in the model's latent representations. Specifically, on MBPP \cite{mbpp} dataset, we obtain erroneous outputs of CodeGen-2B \cite{codegen}, revisions of the outputs, and samples in MBPP. We concatenate the requirements with their code, input them into CodeGen-2B, and extract the hidden representations from the model's final layer. Then, we compute the Euclidean distances within the model's representational space to quantify the disparities between these three elements. The experimental results show that the average distance between the model's erroneous outputs and the dataset's samples is 12.35, whereas the average distance between the erroneous outputs and their revisions is significantly lower, at 6.39. 
\textbf{These experimental results suggest that within the model's representation space, revised codes are closer and similar to the erroneous output codes than the original code samples. This evidence supports our hypothesis of why the error-driven learning method is more efficient from the perspective of model optimization.}

Therefore, our work is determined to explore the use of error-driven learning to achieve a data-efficient adaptation method, aimed at enhancing the performance of LLMs in specific code generation scenarios.
\begin{figure*}[t!]
\centering
\includegraphics[width=1 \textwidth]{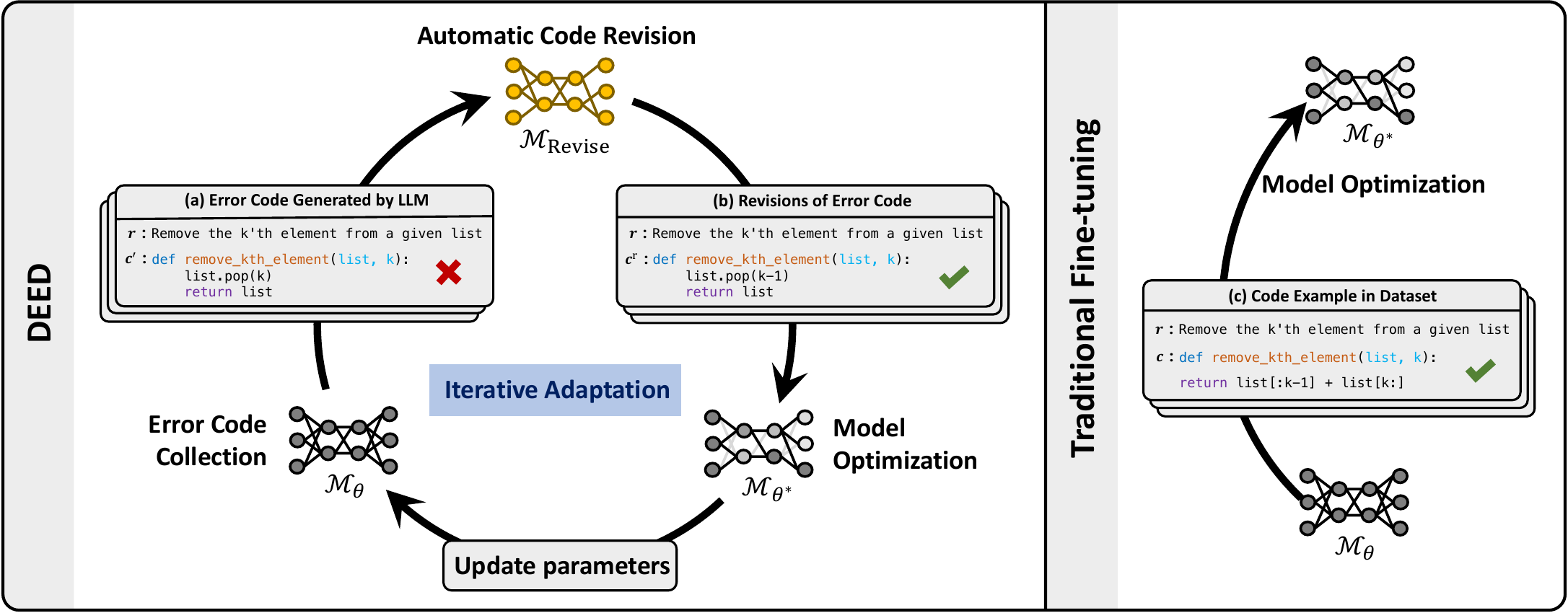}
\caption{An overview of the proposed \ourapproach and its differences from traditional fine-tuning methods.}
\label{method}
\end{figure*}

\section{Methodology}
Given a code generation scenario/domain with a limited-sample training dataset $\mathcal{D}_{train} = \{ (r, c) \}$, where each data pair $(r, c)$ consists of an input requirement $r$ and an associated example of desired output code $c$. For a pre-trained LLM $\mathcal{M}_{\theta}$ with parameter $\theta$, we aim to adapt $\mathcal{M}_{\theta}$ to the specific scenario of $\mathcal{D}_{train}$. 
Inspired by error-driven learning, we propose \ourapproach to achieve data-efficient adaptation of LLMs. \ourapproach consists of the following four steps: Error Code Collection (\S \ref{step1}), Automatic Code Revision (\S \ref{step2}), Model Optimization (\S \ref{step3}), and Iterative Adaptation (\S \ref{step4}). 
LLMs are well adapted after applying \ourapproach, and their subsequent use (inference) remains unchanged compared to direct generation with LLMs, while improving the accuracy of one-time predictions without introducing any additional overhead.
The overview of \ourapproach and its differences from traditional fine-tuning are shown in Figure \ref{method}.

\subsection{Error Code Collection} \label{step1}
In this step, we systematically identify and collect erroneous output of LLMs using testing as criteria.
We employ rejection sampling \cite{casella2004generalized} to draw error code samples from the distribution produced by \(\mathcal{M}_{\theta}\).  
For each requirement $r \in \mathcal{D}_{train}$, we sample 
\begin{equation}
    \label{rejectsample}
    c' \sim \mathcal{M}_{\theta}(r) ~|~ \neg f,
\end{equation}
where we sample multiple times and employ the criterion function \( f \) to determine the retention of the code sample. Specifically, the error code sample \(c'\) is retained when \( f(r, c') = 0 \), where  $f(r, c') = 0$ if the rejection condition is satisfied, otherwise $1$. 

We define \( f \) as a test evaluation function since testing is the criterion for judging the correctness of code in practice:
\begin{equation}\label{func:unit-eval}
   \textsc{TestEval}(r, c') \coloneqq \left\{
	\begin{array}{ll}
             0, & \text{if } c' \text{ fails } S_r, \\
		1, & \text{otherwise},
	\end{array}
\right.
\end{equation}
where \(S_r\) is a suit of test cases under the requirement $r$ and is equipped by code generation datasets. 
When collecting error codes for test failures, we can keep the failed test cases and error messages simultaneously for further error diagnosis.

To gain insights into the propensity of \(\mathcal{M}_{\theta}\) to make certain errors, it is advisable to select error code sample \(c'\) for which the model demonstrates relatively high confidence. Therefore, among multiple error codes collected for the same $r$, we select the one with the highest generation probability, which is determined by the average log-probability of each token in the generated code.

\begin{equation}
    \log P(c'|r) = \frac{1}{|c'|} \sum_i \log P(s_i|s_{<i}, r),
\end{equation}
where $s_i$ is the $i$-th output token, $s_{<i}$ denotes the set of previous tokens, and $|c'|$ indicates the length of the code.

\subsection{Automatic Code Revision} \label{step2} 
\begin{figure}[t!]
\centering
\includegraphics[width=0.5\textwidth]{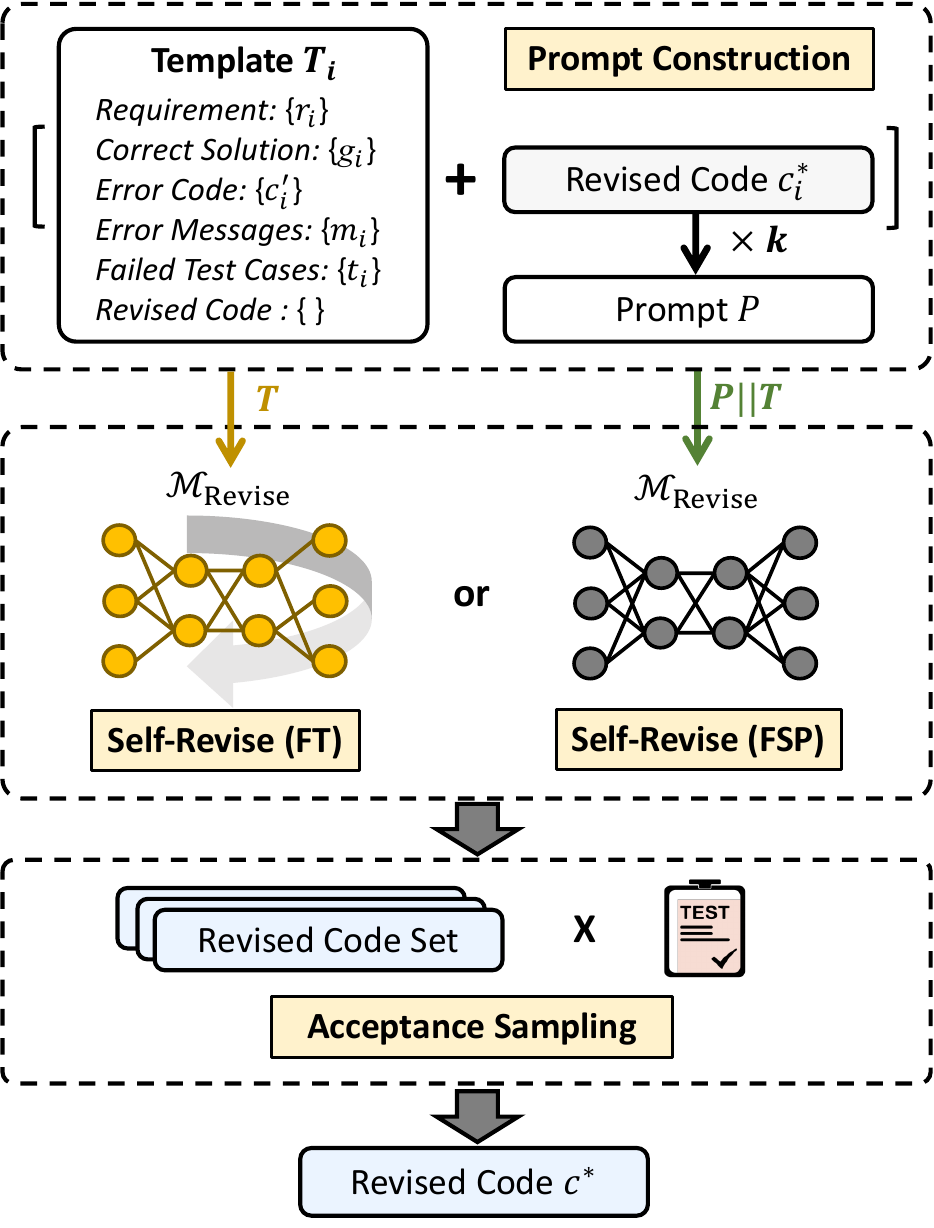}
\caption{Illustration of automatic code revision.}
\label{self_revise}
\end{figure}

In this step, we perform automatic revision for error codes using our \selfrefine method. 
Based on the LLM \(\mathcal{M}_{\theta}\) itself, \selfrefine revises the error code by providing the information in the original dataset and pointing out the error with code execution feedback. 
Our objective is to derive a revised code that fixes the critical bug in the error code. As illustrated by examples (a), (b), and (c) in Figure \ref{method}, although there is already a correct code $(c)$ in the dataset, it may differ significantly from the error code $(a)$, leading to the critical bug being unclear.
The pipeline of automatic code revision is shown in Figure \ref{self_revise}.
Specifically, we leverage the following parts as the input of \selfrefine: 
\begin{itemize}
     \item \textbf{Requirement ($\mathbf{r}$)}: Clarify the requirement that need to be addressed;
     \item \textbf{Correct Solution ($\mathbf{g}$)}: Provide a type of correct solution as a reference to reduce the difficulty of revision. The correct solution used here is the code sample $c$ in the training dataset;
     \item \textbf{Error Code ($\mathbf{c'}$)}: Give the error code that needs to be revised. The error code is generated by \(\mathcal{M}_{\theta}\) under $r$;
     \item \textbf{Error Messages ($\mathbf{m}$) and Failed Test Cases ($\mathbf{t}$)}: Point out the error messages received during execution and the specific test cases where the error code fails, allowing for more focused troubleshooting and revision.
\end{itemize}
These parts are combined as the input of \selfrefine according to the template:
\begin{equation}
        T =  \operatorname{Template}(r, g, c', m, t) 
    \label{T}
\end{equation}
where $\operatorname{Template}$ is shown in Figure \ref{self_revise}.

Following previous work \cite{Self-Edit, Self-collaboration}, we use two settings for \selfrefine, i.e., fine-tuning (FT) and few-shot prompting (FSP), to get \(\mathcal{M}_{\operatorname{Revise}}\) for revising error codes.

\textbf{\selfrefinebf (FT)} entails the process of fine-tuning \(\mathcal{M}_{\theta}\) with a small number of 
data for the purpose of automatic code revision. The training objective is to minimize $L(\theta)$:
\begin{equation}
\label{Loss}
L(\theta) = \sum_{i} l_{ce}(\mathcal{M}_{\theta}(T_i), c^*_i)
\end{equation}
where $l_{ce}$ represents the standard cross-entropy loss, and we update the parameters initialized with \(\mathcal{M}_{\theta}\) to obtain $\mathcal{M}_{\operatorname{Revise}}$ in \selfrefine (FT). 

\textbf{\selfrefinebf (FSP)} adopts the few-shot prompting technique and leverages $k$ examples of $T_i$ and $c^*_i$ to construct the prompt $P$ for aligning \(\mathcal{M}_{\theta}\) to automatic code revision. In \selfrefine (FSP), $\mathcal{M}_{\operatorname{Revise}}(\mathbf{\cdot})$ is defined as $\mathcal{M}_{\theta}(P \ || \ \mathbf{\cdot} \ )$, where $||$ denotes the concatenation operation.

In contrast to the previous error code collection step, for each error code $c'$, we construct $T$ and use acceptance sampling to obtain the revised code $c^*$:
\begin{equation}
    \label{acceptancesampling}
    c^* \sim \mathcal{M}_{\operatorname{Revise}}(T) ~|~ f.
\end{equation}
where $c^*$ is retained if $\textsc{TestEval}(r, c^*) = 1$ in Eq. \eqref{func:unit-eval}, i.e., the revised code $c^*$ passes its test cases. We sample multiple times, and it is sufficient if \(\mathcal{M}_{\operatorname{Revise}}\) could correctly revise the error code once.
To prevent \(\mathcal{M}_{\operatorname{Revise}}\) from simply replicating the provided correct solution \(g\), we exclude the output identical to \(g\). Subsequently, for each requirement $r$, select the version that is most similar to the error code among the remaining code revisions.

\subsection{Model Optimization} \label{step3}
In this step, we employ pairs of the requirement $r$ and its revised code $c^*$ to further fine-tune the model $\mathcal{M}_{\theta}$. This process leads to the enhanced version of $\mathcal{M}_{\theta}$, referred to as $\mathcal{M}_{\theta^*}$, in the specific scenario of dataset $\mathcal{D}_{train}$.

For fine-tuning \cite{bert}, we update all parameter $\theta$ of LLMs as
\begin{equation}
\label{ft}
\theta^* = \arg\min_{\theta} \sum_{(r, c^*)} l_{ce}(\mathcal{M}_{\theta}(r), c^*),
\end{equation}

When the computational resources are insufficient, we employ Low-Rank Adaptation (LoRA) \cite{LORA} to fine-tune LLMs. For a weight matrix $W \in \mathbb{R}^{d \times k}$, LoRA represents its update with a low-rank decomposition:
\begin{equation}
\label{w}
    W + \Delta W = W + \Delta \alpha W_{\operatorname{down}} W_{\operatorname{up}},
\end{equation}
where $\alpha$ is a tunable scalar hyperparameter, $W_{\operatorname{down}} \in \mathbb{R}^{d \times r}$, $W_{\operatorname{up}} \in \mathbb{R}^{r \times k}$, and $r << min(r, k)$. 
In LoRA, we update parameter $\theta^*$ as 
\begin{equation}
    \theta^* = \theta + \Delta\theta, \label{theta}
\end{equation}
\begin{equation}
     \Delta\theta = \arg\min_{\Delta\theta} \sum_{(r, c^*)} l_{ce}(\mathcal{M}_{\theta + \Delta\theta}(r), c^*).\label{delta}
\end{equation}

\subsection{Iterative Adaptation} \label{step4}
The preceding three steps can go through multiple iterations until a certain number of rounds is reached or the revised code no longer increases. 

For $l$-th iteration that $l > 1$, we initialize its initial model $\mathcal{M}_{\theta_{l}}$ as the enhanced model of the previous iteration $\mathcal{M}_{\theta^*_{l-1}}$. Based on $\mathcal{M}_{\theta_{l}}$, we repeat the process in steps of error code collection and automatic code revision to sample error codes $\{c'\}_{l}$ and revised codes $\{c^*\}_{l}$, respectively. 
Inspired by experience replay \cite{experience_replay} in reinforcement learning, we use the union of collected data in each previous iteration $\{(r, c^*)\}_{1:l}$ to stabilize the learning process and improve data utilization efficiency, i.e.,
\begin{equation}
    \label{union}
    \{(r, c^*)\}_{1} \cup \cdots \cup \{(r, c^*)\}_{i} \cdots \cup \{(r, c^*)\}_{l},
\end{equation}
to update parameters in the model optimization step, thereby yielding the enhanced model of the $l$-th iteration $\mathcal{M}_{\theta^*_{l}}$.
At each iteration, the model is trained to convergence. 

This iteration is a step-by-step optimization designed to continuously improve the adaptability of models to the specific scenario. The complete process of \ourapproach is summarized in Algorithm \ref{algorithm}.

\begin{algorithm}[h]
\caption{Pseudocode of \ourapproach.}\label{algorithm}
\small{
\begin{algorithmic}[1]
\REQUIRE{Dataset $\mathcal{D}_{train} = \{(r, c)\}$, initial LLM $\mathcal{M}_{\theta}$.}
\ENSURE{LLM $\mathcal{M}_{\theta^*}$.}\\
\STATE Initial iteration index $l = 0$ and $\mathcal{M}_{\theta_{l+1}} = \mathcal{M}_{\theta}$. \\[5pt]
\STATE \textcolor{gray}{\textit{\# \textbf{Iterative Adaptation}}} 
\REPEAT
\STATE Update $l = l+1$.\\[5pt]

\STATE \textcolor{gray}{\textit{\# \textbf{Error Code Collection}}} 
    
   \STATE Perform rejection sampling to collect error codes $\{c'\}_{l}$ based on $\mathcal{M}_{\theta_l}$ via Eq. \eqref{rejectsample} and \eqref{func:unit-eval}. \\[5pt]
   
\STATE \textcolor{gray}{\textit{\# \textbf{Automatic Code Revision}}} 
    
   \STATE Perform acceptance sampling to collect revised codes $\{c^*\}_{l}$ based on $\mathcal{M}_{\theta_l}$ and \selfrefine via Eq. \eqref{func:unit-eval}, \eqref{T}, and \eqref{acceptancesampling}.

   \STATE Calculate the union of $\{(r, c^*)\}_{1:l}$ via Eq. \eqref{union}. \\[5pt]
   
\STATE \textcolor{gray}{\textit{\# \textbf{Model Optimization}}} 
    
   \STATE Fine-tune $\mathcal{M}_{\theta_l}$ to yield $\mathcal{M}^*_{\theta_l}$ via Eq. \eqref{ft} if the computational resources are sufficient, otherwise via Eq. \eqref{w}, \eqref{theta}, and \eqref{delta}.
   \STATE Update $\mathcal{M}_{\theta_{l+1}} = \mathcal{M}_{\theta^*_{l}}$.

\UNTIL{End condition is satisfied}
\RETURN{$\mathcal{M}_{\theta^*_{l}}$}
\end{algorithmic}
}
\end{algorithm}
\section{Evaluation Setup}
We present extensive experiments that span five representative code generation datasets, two fine-tuning settings, and four different LLMs of varying sizes or series. We aim to investigate six research questions: 
\begin{itemize}
\item \textbf{RQ1: How does \ourapproachbf perform compared to the mainstream adaptation approaches?} This question aims to investigate the superiority of our method compared to existing approaches in data-scarce scenarios.

\item \textbf{RQ2: How does \ourapproachbf perform when applied to various LLMs?}
This question explores the usefulness of \ourapproach across different LLMs, assessing its applicability and performance consistency in varying LLMs.

\item \textbf{RQ3: What kind of training sample has the better training effect?}
This question helps us understand the promotion of training efficiency by training with the revision of LLMs' erroneous outputs.

\item \textbf{RQ4: How does the number of iterations affect the effectiveness of \ourapproachbf?}
This question explores the changes in LLMs' performance during the process of iterative optimization, to demonstrate the necessity of iterative adaptation in our method.

\item \textbf{RQ5: What is the impact of implementing the automatic code revision component of \ourapproachbf in conjunction with alternative LLMs?}
We choose to use LLMs themselves for revision, termed \selfrefine. This question explores the effects of using other LLMs for revision.

\item \textbf{RQ6: How does each input component of \selfrefinebf contribute to the effectiveness?}
Since \selfrefine utilizes components such as correct solutions, failed test cases, and error messages, this question aims to analyze the impact of these components on its performance.
\end{itemize}

\subsection{Datasets}

We use five public code generation datasets to simulate the specific scenario with limited data and apply \ourapproach to each dataset to evaluate its effectiveness. Specifically,

\begin{itemize}
    \item \textbf{HumanEval} \cite{codex} is a widely-used code generation benchmark, containing 164 handwritten programming problems, proposed by OpenAI. Each programming problem includes a function signature, a natural language description, use cases, a correct solution in Python, and several test cases.
    \item \textbf{MBPP} \cite{mbpp} contains crowd-sourced Python programming problems. We use the version in the work \cite{human_feedback}, which consists of 276 problems and some generated error codes alongside their human-revised counterparts, thus facilitating subsequent experiments. 
    \item \textbf{HumanEval-ET} and \textbf{MBPP-ET} \citep{CodeScore} are extended versions of MBPP and HumanEval, respectively, where each includes over 100 additional test cases per problem. This updated version enhances the soundness of code evaluation compared to the original benchmarks.
    \item \textbf{DataScience} \cite{DS-pandas} comprises 291 data science problems utilizing Pandas libraries. This dataset can evaluate the ability of LLMs to utilize specific data-analysis libraries for code generation. 
\end{itemize}

We sample $\min(200, 40\%*\mathcal{D})$ problems from the datasets as $\mathcal{D}_{{train}}$, while the remaining problems serve as $\mathcal{D}_{{test}}$.

\subsection{Implementation Details}

We use a single A6000 GPU to conduct all experiments. We select CodeGen-2B \cite{codegen} as our base model by default, which is a well-known open-source LLM for code and is suitable for full fine-tuning within our computational resource constraints. $\mathcal{M}_{\theta}$ is initialized to our base model, and $\mathcal{M}_{\operatorname{Revise}}$ is derived from $\mathcal{M}_{\theta}$ though \selfrefine (\S\ref{step2}). In fine-tuning setting, $\mathcal{M}_{\operatorname{Revise}}$ only needs to be trained at the beginning and then remains unchanged for subsequent operations.

For full parameter fine-tuning, i.e., Fine-tuning (Full) \cite{bert}, we use the AdamW optimizer \cite{AdamW}, with hyperparameters $\beta_1 = 0.9$ and $\beta_2 = 0.9$, accompanied by a linear learning rate schedule. The initial learning rate is set to 5e-6, with a batch size of 1 and gradient accumulation of 32 steps for training across 10 epochs. 
For parameter-efficient fine-tuning, i.e., Fine-tuning (LoRA) \cite{LORA}, the learning rate is set to 2e-4. Additionally, the rank $r$ is adjusted to 128, and the scaling factor $\alpha$ is set at 8. All other hyperparameters remain aligned with Fine-tuning (Full). For few-shot prompting \cite{gpt-3}, we set the number of examples in prompt to 4.
All baselines in the experiments use consistent settings.

In the error code collection step (\S\ref{step1}) and the automatic code revision step (\S\ref{step2}), we use temperature \cite{tempSample1,tempSample2} sampling to generate multiple samples: 5 samples in the former and 30 in the latter, with the temperature set to 0.8. To obtain the final revised code in the automatic code revision step, we choose one of the revised codes exhibiting the minimum Levenshtein distance \cite{levenshtein1966binary} to the error code. The number of iterative adaptations is set to 2. 
The maximum generation length is uniformly limited to 1024 tokens.

\subsection{Metrics}
Following the practice of real software development which utilizes testing for evaluation \cite{lifecycle_models, abrahamsson2002agile}, we employ the Pass@k \cite{alphacode} metric to measure the functional correctness of the generated code by executing test cases.
We use the unbiased version \cite{codex} of Pass@k, where $n>=k$ samples are generated for each problem, count the number of correct samples $c<=n$ which pass test cases and calculate the following estimator,

\begin{equation}
\label{Pass@k}
    \footnotesize{\operatorname{Pass@k} = \mathop{\mathbb{E}}\limits_{\operatorname{Problems}}\begin{bmatrix}1-\frac{\begin{pmatrix}n-c\\k\end{pmatrix}}{\begin{pmatrix}n\\k\end{pmatrix}}\end{bmatrix}.}
\end{equation}

For automatic code revision, we add the pass@any metric which refers to the percentage of tasks for which the model generates at least one correct code that passes all test cases.

In the final evaluation of this paper, we set the temperature to 0.8 and generate $n = 50$ samples, which are then used to calculate unbiased Pass@k \cite{codex} via Eq. \eqref{Pass@k} in all experiments. All evaluation results are averaged over five test runs.

\section{Results}

\subsection{RQ1: Comparison of \ourapproachbf and the mainstream adaptation approaches}

\label{RQ1}
\noindent \textbf{Baselines.}
In this section, we evaluate the effectiveness of \ourapproach by comparing it against four mainstream approaches used as baselines:
\begin{itemize}
    \item \textbf{Direct Generation}: We evaluate the LLM directly to demonstrate the performance of the original LLM.
    \item \textbf{Fine-tuning (Full)}: We employ full-parameter fine-tuning for  the LLM on \(\mathcal{D}_{train}\).
    \item \textbf{Fine-tuning (LoRA)}: We fine-tune the LLM on \(\mathcal{D}_{train}\) using LoRA \cite{LORA}, which greatly reduces the number of trainable parameters.
    \item \textbf{Few-shot Prompting}: We use 4-shot prompt \cite{gpt-3} to align LLMs with the input-output format of \(\mathcal{D}_{train}\), where 4 examples in prompt are randomly selected from \(\mathcal{D}_{train}\).
    \item \textbf{Self-refine} \cite{Self-Refine} and \textbf{Self-debug} \cite{debug} iteratively refine the generated code through prompting techniques. The number of iterations is set to 2.
\end{itemize}
Considering the baselines involve full-parameter fine-tuning, CodeGen-2B is uniformly selected as the base model in this experiment.
For \ourapproach, we use 30\% data of \(\mathcal{D}_{train}\) for \selfrefine (FT)\footnote{In addition to MBPP dataset, for the other two datasets (i.e., HumanEval and DataScience), we generate one error code per sample in a subset comprising 30\% of the training set, using CodeGen-2B. Subsequently, authors collaboratively apply the minimal necessary revisions to correct these error codes.}, while the remaining 70\% data of \(\mathcal{D}_{train}\) is employed for model optimizing, where we use full-parameter fine-tuning.

\begin{table}[h!]
\caption{Pass@k (\%) of \ourapproach and baselines on HumanEval, MBPP, and DataScience datasets. The values in brackets represent results on the extended version of the datasets, and the teal number after $\uparrow$ denotes the relative improvement of \ourapproach over the second-highest score.
}
\centering
{
\begin{tabular}{clccc}
\toprule
\multicolumn{1}{l}{\multirow{1}{*}{Datasets}} & 
  \multicolumn{1}{l}{Method} &
  \multicolumn{1}{c}{Pass@1} &
  \multicolumn{1}{c}{Pass@5} &
  \multicolumn{1}{c}{Pass@10} \\

  \midrule 
  \multicolumn{1}{l}{\multirow{6}{*}{HumanEval}} & Direct Generation     & 24.8\%          & 44.7\%          & 51.8\%          \\
                   & Fine-tuning (Full)    & 29.8\%          & 47.9\%          & 56.4\%          \\
                   & Fine-tuning (LoRA)    & 27.4\%          & 46.9\%          & 53.9\%          \\
                   & Few-shot Prompting    & 25.2\%          & 45.8\%          & 53.1\%          \\
                   & Self-Refine           & 25.3\%          & 45.2\%          & 51.9\%          \\
                   & Self-Debug            & 26.4\%          & 46.4\%          & 54.2\%          \\\cdashline{2-5}
                   & \ourapproach & $\textbf{38.6\%} \ \ (\textcolor{teal}{\uparrow 29.5\%})$ & \textbf{54.7\%} & \textbf{62.2\%} \\ 
   \midrule 
  \multicolumn{1}{l}{\multirow{6}{*}{HumanEval-ET}} & Direct Generation     & 17.0\%          & 30.8\%          & 36.3\%          \\
                   & Fine-tuning (Full)    & 21.5\%          & 34.2\%          & 41.5\%          \\
                   & Fine-tuning (LoRA)    & 20.3\%          & 33.7\%          & 39.9\%          \\
                   & Few-shot Prompting    & 19.1\%          & 33.5\%          & 39.1\%          \\
                   & Self-Refine           & 17.5\%          & 32.6\%          & 38.2\%          \\
                   & Self-Debug            & 19.2\%          & 33.6\%          & 41.0\%          \\
  \cdashline{2-5}
   \multicolumn{1}{l}{} & \ourapproach  & $\textbf{28.6\%} \ \ (\textcolor{teal}{\uparrow 33.0\%})$ & \textbf{39.3\%} & \textbf{46.5\%} \\ 
  \midrule 
  \multicolumn{1}{l}{\multirow{6}{*}{MBPP}} & Direct Generation     & 15.6\%          & 31.4\%          & 40.2\%          \\
              & Fine-tuning (Full)    & 25.8\%          & 45.2\%          & 57.6\%          \\
              & Fine-tuning (LoRA)    & 19.8\%          & 39.8\%          & 55.2\%          \\
              & Few-shot Prompting    & 24.4\%          & 38.0\%          & 49.4\%          \\
              & Self-Refine           & 25.6\%          & 38.8\%          & 50.2\%          \\
              & Self-Debug            & 20.2\%          & 34.5\%          & 40.6\%          \\ 
  \cdashline{2-5}
    & \ourapproach  & $\textbf{32.8\%} \ \ (\textcolor{teal}{\uparrow 27.1\%})$ & \textbf{46.8\%} & \textbf{64.0\%}\\ 
   \midrule 
   \multicolumn{1}{l}{\multirow{6}{*}{MBPP-ET}} & Direct Generation     & 10.6\%          & 21.7\%          & 28.2\%          \\
                   & Fine-tuning (Full)    & 18.1\%          & 33.1\%          & 43.7\%          \\
                   & Fine-tuning (LoRA)    & 15.1\%          & 32.5\%          & 42.4\%          \\
                   & Few-shot Prompting    & 17.5\%          & 27.3\%          & 36.9\%          \\
                   & Self-Refine           & 17.8\%          & 27.6\%          & 37.2\%          \\
                   & Self-Debug            & 14.1\%          & 25.7\%          & 29.6\%          \\
  \cdashline{2-5}
 & \ourapproach  & $\textbf{24.9\%} \ \ (\textcolor{teal}{\uparrow 37.6\%})$ & \textbf{36.3\%} & \textbf{49.3\%} \\ 
  \midrule 
  \multicolumn{1}{l}{\multirow{6}{*}{DataScience}} & Direct Generation   & 0.8\% & 3.1\% & 5.6\% \\
  \multicolumn{1}{l}{} & Fine-tuning (Full)  & 2.6\% & 6.5\% & 9.6\% \\
  \multicolumn{1}{l}{} & Fine-tuning (LoRA) & 2.2\% & 6.0\% & 8.9\% \\
  \multicolumn{1}{l}{} & Few-shot Prompting & 1.9\% & 4.5\% & 5.7\%\\
  \multicolumn{1}{l}{} & Self-Refine & 2.1\% & 4.7\% & 5.8\% \\ 
  \multicolumn{1}{l}{} & Self-Debug & 1.2\% & 2.8\% & 4.1\% \\ 
  \cdashline{2-5}
  \multicolumn{1}{l}{} & \ourapproach  & $\textbf{5.3\%} \ \ (\textcolor{teal}{\uparrow 103.8\%})$ & \textbf{9.5\%} & \textbf{12.3\%} \\  
\bottomrule
\end{tabular}
}\label{Baselines}
\end{table}

\noindent \textbf{Results.}
We conducted experiments on five public datasets, i.e., MBPP, HumanEval, and DataScience. 
The experimental results are summarized in Table \ref{Baselines}. This comparison yielded four insightful observations:
\textbf{1) Significant superiority of \ourapproachbf}: Our proposed \ourapproach performs significantly better than the other six baselines on the five datasets. Notably, \ourapproach exhibits significant relative improvements of 29.5\%, 33.0\%, 27.1\%, 37.6\%, and 103.8\%, respectively, when compared to the best-performing baseline Fine-tuning (Full). 
Self-refine and Self-debug underperform on small LLMs like codegen-2B. Self-debug excels over Self-refine only on the HumanEval dataset, where public test cases and results are available. On other datasets, Self-debug relies on the LLM's generated code explanations and feedback.
Moreover, we find that \ourapproach not only surpasses Self-refine and Self-debug in terms of performance but also in cost. These methods have a significant disadvantage in cost as they require iterative refinements for each sample, leading to increased time and token consumption. Their cost is directly proportional to the number of sampling and the number of iterations, while \ourapproach is free of these two factors during inference.
\textbf{2) Worst performance of Direct Generation}: The performance of Direct Generation is significantly lower than Fine-tuning (Full), Fine-tuning (LoRA), and Prompt baselines. This result suggests that directly applying LLMs for evaluation may be less suitable for specific scenarios, resulting in performance differences. 
\textbf{3) Fine-tuning (LoRA) is less effective than Fine-tuning (Full)}: Although Fine-tuning (LoRA) offers the advantage of reduced computational resource requirements for fine-tuning LLMs, it trades off the performance.
\textbf{4) Less improvement of Few-shot Prompting}: 
Few-shot prompting is the most commonly used prompting technique, but its main limitation lies in its difficulty in imparting new knowledge or developing new capabilities in the model. It primarily assists the model in adjusting its outputs to better align with expected results, therefore its adaptability is limited.

\vspace{6pt}

\RS{1}{In code generation scenarios with limited training data, \ourapproach exhibits improvements compared to the mainstream adaptation approaches, achieving relative improvements between 27.2\% and 103.9\%. }

\subsection{RQ2: \ourapproachbf with Different LLMs} \label{RQ2}
\noindent \textbf{Baselines.}
We use the following types of well-known LLMs to perform \ourapproach, including 
\begin{itemize}
    \item \textbf{CodeGen-2B and CodeGen-6B} \cite{codegen} are code LLMs trained on natural language and programming data for conversation-based program synthesis. 
    
    \item \textbf{Llama-7B} \cite{llama2} is a general-purpose foundational LLM that has been trained on diverse data to handle a wide range of tasks, which is developed by Meta. 

    \item \textbf{CodeLlama-7B} \cite{CodeLlama}
    is an open foundational LLM for code generation tasks, derived from continuous training and fine-tuning based on Llama architecture \cite{llama2}.
\end{itemize}
Among them, CodeGen-2B uses full fine-tuning, and the remaining LLMs use parameter-efficient fine-tuning with LoRA. 
Each LLM has the same baselines as RQ1 (\S\ref{RQ1}), i.e., \textbf{Direct Generation}, \textbf{Fine-tuning}, and \textbf{Few-shot Prompting}.

\begin{table}[h!]
\caption{Pass@k (\%) of \ourapproach and baselines with different LLMs, and the teal number after $\uparrow$ denotes the relative improvement of \ourapproach over Fine-tuning.}
\centering
{
\begin{tabular}{clccc}
\toprule
\multicolumn{1}{l}{\multirow{1}{*}{Models}} & 
  \multicolumn{1}{l}{Method} &
  \multicolumn{1}{c}{Pass@1} &
  \multicolumn{1}{c}{Pass@5} &
  \multicolumn{1}{c}{Pass@10} \\
  \midrule 
  \multicolumn{1}{l}{\multirow{4}{*}{CodeGen-2B}} & Direct Generation & 15.6\% & 31.4\% & 40.2\% \\
  \multicolumn{1}{l}{} & Fine-tuning (Full)   & 25.8\% & 45.2\% & 57.6\% \\
  \multicolumn{1}{l}{} & Few-shot Prompting & 24.4\% & 38.0\% & 49.4\% \\ 
  \cdashline{2-5}
  \multicolumn{1}{l}{} & \ourapproach (Full)   & $ \textbf{32.8\%} \ \ (\textcolor{teal}{\uparrow 27.1\%})$ & \textbf{46.8\%} & \textbf{64.0\%} \\ 
  \midrule 
  \multicolumn{1}{l}{\multirow{4}{*}{CodeGen-6B}} & Direct Generation   & 19.6\% & 40.2\% & 60.8\% \\
  \multicolumn{1}{l}{} & Fine-tuning (LoRA)    & 26.6\% & 46.8\% & 63.0\% \\
  \multicolumn{1}{l}{} & Few-shot Prompting & 26.2\% & 45.2\% & 60.2\% \\
  \cdashline{2-5}
  \multicolumn{1}{l}{} & \ourapproach (LoRA)  & $ \textbf{33.4\%} \ \ (\textcolor{teal}{\uparrow 25.6\%})$ & \textbf{47.4\%} & \textbf{67.6\%} \\
  \midrule 
  \multicolumn{1}{l}{\multirow{4}{*}{Llama-7B}} & Direct Generation  & 13.4\% & 29.8\% & 37.4\% \\
  \multicolumn{1}{l}{} & Fine-tuning (LoRA)   & 15.2\% & 27.4\% & 34.0\% \\
  \multicolumn{1}{l}{} & Few-shot Prompting & 16.6\% & 26.2\% & 33.8\%\\
  \cdashline{2-5}
  \multicolumn{1}{l}{} & \ourapproach (LoRA) & $ \textbf{22.0\%} \ \ (\textcolor{teal}{\uparrow 32.5\%})$ & \textbf{30.4\%} & \textbf{40.8\%} \\ 
  \midrule 
  \multicolumn{1}{l}{\multirow{4}{*}{CodeLlama-7B}} & Direct Generation   & 20.4\% & 43.8\% & 52.8\% \\
  \multicolumn{1}{l}{} & Fine-tuning (LoRA)    & 19.9\% & 42.4\% & 53.2\% \\
  \multicolumn{1}{l}{} & Few-shot Prompting & 27.8\% & 46.6\% & 64.8\% \\
  \cdashline{2-5}
  \multicolumn{1}{l}{} & \ourapproach (LoRA) & $ \textbf{34.8\%} \ \ (\textcolor{teal}{\uparrow 25.2\%})$ & \textbf{49.2\%} & \textbf{65.8\%} \\   
\bottomrule
\end{tabular}
}\label{sample}
\end{table}

\noindent \textbf{Results.} The results of applying \ourapproach to different LLMs are shown in Table \ref{sample}. The results demonstrate that \ourapproach consistently achieves significant improvements across all these LLMs, outperforming the Direct Generation, Fine-tuning, and Few-shot Prompting baselines. Moreover, we also find that the performances of Code LLMs are better than pure LLMs, and there is a clear trend indicating that as the number of parameters in the LLMs increases, the efficacy of the Direct Generation and Few-shot Prompting approaches increases steadily. In contrast, the effectiveness of \ourapproach on these LLMs is also continuously improving.  

\vspace{6pt}

\RS{2}{\ourapproach consistently enhances performance across all LLMs, outperforming Direct Generation, Fine-tuning, and Few-shot Prompting baselines, showcasing the applicability of \ourapproach.}

\subsection{RQ3: The Effect of Training Sample Variants} \label{RQ3}

\noindent \textbf{Baselines.}
We investigate the influence of different training data on the final adapted model $\mathcal{M}_{\theta^*}$ to validate the effectiveness of using revisions of model’s erroneous output for training. The different variants of training data include:

\begin{itemize}
    \item \textbf{W/o Training}: Direct generation without without any training data.
    
    \item \textbf{Raw $\boldsymbol{\mathcal{D}_{train}}$}: All samples in $\mathcal{D}_{train}$, which are the raw requirement and code sample pairs in the dataset.
    
    \item $\boldsymbol{\mathcal{D}_{train} \cap~}$ \ourapproachbf: 
    The samples of the same problem as \ourapproach in $\mathcal{D}_{train}$
    
    \item \ourapproachbf $ \boldsymbol{\cup~\mathcal{D}_{train}}$: 
    Include not only samples of problems obtained through \selfrefine, but also samples of other problems in $\mathcal{D}_{train}$.
    
    \item \textbf{Human-revised $\boldsymbol{\mathcal{D}_{train}}$}: Samples obtained from human revision, which are the pairs of requirement and revised code of LLM's erroneous output.

    \item \textbf{\ourapproachbf}: Samples obtained through \selfrefine, which are the pairs of requirement and revised code of LLM's erroneous output.
\end{itemize}

\begin{table}[h!]
\caption{Comparison of the effect of different training data variants.}
\centering
{
\begin{tabular}{lcccc}
\toprule
 Variants & Pass@1 & Pass@5 & Pass@10 \\ \hline
W/o Training   & 15.6\% & 31.4\% & 40.2\%\\
Raw $\mathcal{D}_{train}$ & 25.8\% & 45.2\% & 57.6\%\\
 $\mathcal{D}_{train} \cap~$ \ourapproach & 22.4\% & 33.8\% & 42.8\% \\
\ourapproach $ \cup~\mathcal{D}_{train}$ & 29.2\% & 44.2\% & 58.0\% \\
Human-revised $\mathcal{D}_{train}$ & 28.0\% & 46.2\% & 59.8\%\\
\hdashline
\ourapproach  & 32.8\% & 46.8\% & 64.0\% \\
\bottomrule
\end{tabular}
}\label{ablation}
\end{table}

\noindent \textbf{Results.}
As shown in Table \ref{ablation}, we discover that: \textbf{1) \ourapproachbf exceeds Raw $\boldsymbol{\mathcal{D}_{train}}$, despite Raw $\boldsymbol{\mathcal{D}_{train}}$ having more training data.}
This proves that training using revisions produced by \selfrefine is more efficient compared to using samples in the dataset.
\textbf{2) The effect of $\boldsymbol{\mathcal{D}_{train} \cap~}$ \ourapproachbf is comparatively weaker}, which reveals that \ourapproach is not simply improved by selecting better problems. \textbf{3) \ourapproachbf $ \boldsymbol{\cup~\mathcal{D}_{train}}$ is not as effective as \ourapproachbf}, which shows that some samples in $\mathcal{D}_{train}$ have negative effects for training. \textbf{4) The performance of \ourapproachbf surpasses that of the Human-revised $\boldsymbol{\mathcal{D}_{train}}$.} This finding may be attributed to a disconnect between the revision made by humans and the model's learning expectations. While human revisions are applied to all code samples in $\mathcal{D}_{train}$, some samples may inherently be challenging for the current model. As such, forced learning from these samples may have counterproductive effects, highlighting a potential limitation in human-revised $\mathcal{D}{train}$.

\vspace{6pt}

\RS{3}{The revisions of error code through \selfrefine surpass other training sample variants, including both samples in the dataset and the samples revised by humans, in terms of training efficiency and effectiveness.}

\subsection{RQ4: The Effect of Iterations} \label{RQ4}
\noindent \textbf{Baselines.}
We study the effect of iterations on \ourapproach. We analyze the progression of \ourapproach's effectiveness across different iterations, starting from 0 iterations (i.e., generated directly with LLMs) and extending to one, and up to four iterations.

\begin{table}[h!]
\caption{Performance of \ourapproach with the different number of iterations, where NRC indicates the Number of Revised Codes added in each iteration, and Loss is calculated at the end of training in each iteration.}
\centering
{
\begin{tabular}{cccccc}
\toprule
Iterations & Pass@1 & Pass@5 & Pass@10 & NRC & Loss \\ \midrule
0 & 15.6\% & 31.4\% & 40.2\% & - & - \\
1 & 31.6\% & 46.3\% & 60.6\% & 31 (+31) & 0.0891\\
\rowcolor[gray]{0.9}
2 & 32.8\% & 46.8\% & 64.0\% & 41 (+10) & 0.0375\\
3 & 33.0\% & 46.7\% & 62.6\% & 43 (+2) & 0.0094 \\
4 & 33.2\% & 47.1\% & 64.0\% & 44 (+1) & 0.0056 \\
\bottomrule
\end{tabular}
}\label{iteration}
\end{table}

\noindent \textbf{Results.}
We conduct this experiment on MBPP dataset, and its results are displayed in Table \ref{iteration}. From the results, we can observe a trend: as the number of iteration rounds increases, the performance of \ourapproach first shows an increasing trend and then gradually stabilizes. The amount of revised code in each iteration is also increasing, indicating that errors are continuously discovered, corrected, and learned. Additionally, the results show a consistent reduction in loss as the number of iterations increases, which affirms the training stability of \ourapproach.
Considering that Pass@10 has oscillations from the 2nd iteration to the 4th iteration, we choose to end after the second iteration as the final performance of \ourapproach.

\vspace{6pt}

\RS{4}{As iteration rounds increase, the performance of \ourapproach initially improves and then stabilizes in Pass@1, and over 98\% performance can be achieved in two iteration.}

\subsection{RQ5: Automatic Code Revision Based on Other Models} \label{RQ5}
\noindent \textbf{Baselines.}
We evaluate the performance of automatic code revision and the impact on the final model $\mathcal{M}_{\theta^*}$, which is obtained through \ourapproach, when using alternative LLMs to substitute the base model as $\mathcal{M}_{\operatorname{Revise}}$. The base model and alternative LLMs are as follows:
\begin{itemize}
    \item \textbf{Base Model}: 
    We use CodeGen-2B \cite{codegen} as the base model for automatic code revision, i.e., \selfrefine.
    \item  \textbf{CodeGen-6B} \cite{codegen}: A variant in the same series as our base model but with a larger parameter count.
    \item \textbf{Llama-7B} \cite{llama2}: A LLM with different training data and architectures of the base model.
    \item \textbf{CodeLlama-7B} \cite{CodeLlama}: A code LLM with different training data and architectures of the base model. 
    \item \textbf{ChatGPT} \cite{ChatGPT}: A powerful AI chatbot developed based on LLMs. Since ChatGPT is closed-source and cannot be fine-tuned, we employ ChatGPT for \selfrefine (FSP).
    \item \textbf{GPT-3.5-turbo} \cite{openai_gpt35_turbo} is an instruction-tuned LLM developed by OpenAI, optimized for dialogue and efficient inference.
\end{itemize}
In this experiment, we obtain $\mathcal{M}_{\operatorname{Revise}}$ in both fine-tuning and few-shot prompting settings for comparison, and $\mathcal{M}_{\theta^*}$ is consistently fixed as the base model.

\begin{table}[h!]
\caption{Comparison of automatic code revisions based on different LLMs and settings as well as their impact on the final model, where $\mathcal{M}_{\operatorname{Revise}}$ represents the \selfrefine model that used during the training and $\mathcal{M}_{\theta^*}$ represents the final model that used during the inference. $\mathcal{M}_{\operatorname{Revise}}$ is reported the raw results on the $70\%*\mathcal{D}_{train}$ part} and $\mathcal{M}_{\theta^*}$ is fine-tuned with filtered results as described in \S\ref{step2}.
\centering
{
\begin{tabular}{lccccc}
\toprule
 \multirow{2}{*}{Method} & \multicolumn{3}{c}{\cellcolor{gray!10} $\mathcal{M}_{\operatorname{Revise}}$} & \multicolumn{2}{c}{\cellcolor{gray!20}  $\mathcal{M}_{\theta^*}$} \\ \cmidrule(r){2-4} \cmidrule(r){5-6}
 & Pass@1 & Pass@10 & Pass@any & Pass@1 & Pass@10\\ \midrule
 
\multicolumn{6}{l}{\textbf{Few-shot Prompting}}\\
CodeGen-6B & 19.4\% & 60.1\% & 70.8\% & 26.8\% & 59.0\% \\ 
Llama-7B & 23.5\% & 67.7\% & 81.9\% & 20.8\% & 54.2\% \\
CodeLlama-7B & 20.2\% & 64.9\% & 75.0\% & 25.2\% & 59.6\% \\
ChatGPT & 61.4\% & 87.3\% & 92.1\% & 27.0\% & 62.4\% \\
GPT-3.5-turbo & 59.4\% & 69.5\% & 72.7\% & 29.0\% & 51.0\% \\
\hdashline
\makecell[l]{Base Model \\ (\selfrefine (FSP))}& 18.9\% & 57.1\% & 69.4\% & 26.2\% & 58.2\% \\
\midrule
\textbf{Fine-tuning}\\
CodeGen-6B & 5.0\% & 20.3\% & 26.6\% & 29.4\% & 64.2\% \\
Llama-7B & 2.7\% & 8.5\% & 12.6\% & 23.2\% & 58.4\%\\
CodeLlama-7B & 5.1\% & 21.0\% & 34.6\% & 24.0\% & 60.2\% \\
\hdashline
\makecell[l]{Base Model \\ (\selfrefine (FT))} & 3.9\% & 18.9\% & 24.6\% & \textbf{32.8\%} & \textbf{64.0\%}\\
\bottomrule
\end{tabular}
}\label{variants}
\end{table}

\noindent \textbf{Results.}
Table \ref{variants} illustrates the experimental results of automatic code revision based on different models, and we can observe that: \textbf{1) \selfrefinebf (FT) employing the same model as the base model yields the best performance of $\mathbf{\mathcal{M}_{\theta^*}}$.} For baselines using other LLMs in fine-tuning, CodeLlama exhibits superior performance in terms of Pass@k in $\mathcal{M}_{\operatorname{Revise}}$, but its final effectiveness is somewhat compromised. This limitation is attributed to the divergence in training data and architectural frameworks between CodeLlama and the base model, leading to inconsistencies in the revised code with the base model's expectations. In contrast, CodeGen-6B, which is the same series of the base model with a large parameter, demonstrates slightly lower Pass@k in $\mathcal{M}_{\operatorname{Revise}}$ than CodeLlama but still achieves commendable results for $\mathcal{M}_{\theta^*}$. \textbf{2) Although the Pass@k of \selfrefinebf (FSP) is higher than \selfrefinebf (FT) in $\mathbf{\mathcal{M}_{\operatorname{Revise}}}$, it does not perform as well on the ultimate $\mathbf{\mathcal{M}_{\theta^*}}$.} We find this discrepancy may be due to the \selfrefine (FSP)'s tendency to learn superficial forms, i.e., it often resorts to directly copying code from the correct solution provided in the prompt, even when explicitly instructed not to in the prompt. 
Using ChatGPT as $\mathcal{M}_{\operatorname{Revise}}$ results in substantially higher Pass@k compared to using the base model, does not significantly enhance the final model $\mathcal{M}_{\theta^*}$. 

To understand why the ChatGPT-based revision underperforms the self-revision model, despite providing more training data, we identify two reasons: 1) We manually check the revised data and find that although the quantity of revisions increased, ChatGPT sometimes disregards the instruction to make "minimal necessary revisions." Instead, it tends to reference the ground-truth correct code, resulting in substantial changes rather than minimal fixes. We conduct a further experiment with OpenAI's GPT-3.5-turbo, and the result is shown in Table \ref{variants}. While this issue is less severe and its final performance is better than that of the revision using ChatGPT, it still did not surpass the Self-Revise (FT). 
2) We believe a second reason is the significant capability gap between ChatGPT and our base model (CodeGen-2B). ChatGPT is a much more powerful model, and some of the revision knowledge it provides may not be effectively understood and absorbed by CodeGen-2B. Instead, the base model may simply memorize the patterns, which does not improve its generalization ability. We conclude that this phenomenon highlights a crucial point: the suitability and quality of data are more important than its quantity.

\begin{figure*}[th!]
\centering
\includegraphics[width=1.02\textwidth]{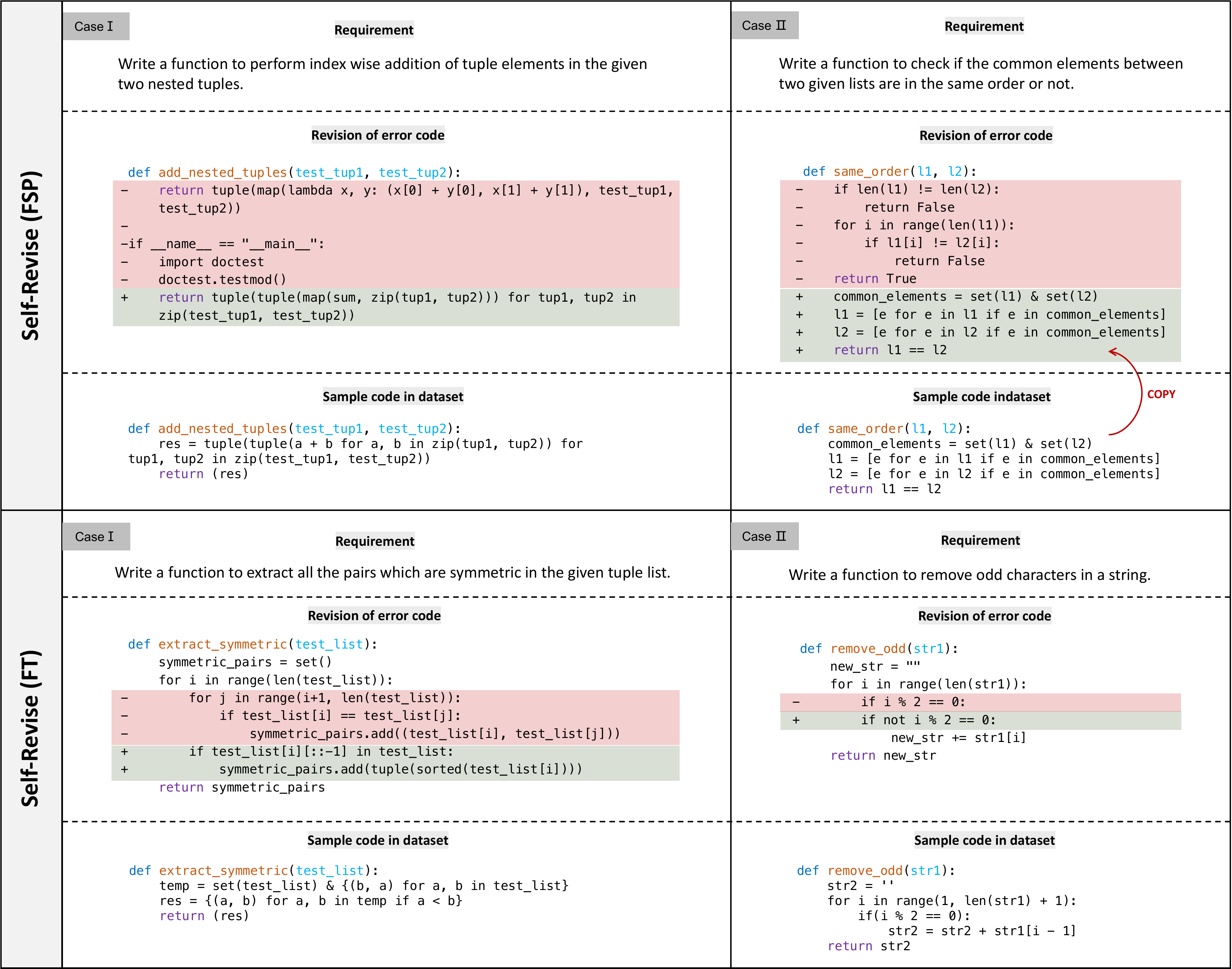}
\caption{Cases for two settings of \selfrefine, where ``-'' and ``+'' respectively indicate lines of code before and after revision.}
\label{cases}
\end{figure*}

\noindent \textbf{Qualitative Examples.}
We use the case study to qualitatively assess the effectiveness of automatic code revision (\S\ref{step2}), i.e., \selfrefine (FSP) and \selfrefine (FT) employed by \ourapproach, examples of which are presented in Figure \ref{cases}. 
Upon manual inspection of the outcomes produced by \selfrefine (FSP), two prevalent modification patterns are identified. First, the removal of redundant code is a common alteration. This includes the deletion of unnecessary blocks such as ``if name == `main'\,'' and other test codes, which are often extraneous in the context of the desired output. Second, \selfrefine (FSP) exhibits a tendency to directly copy correct code samples from the prompt. In contrast, \selfrefine (FT) is capable of making minimal yet effective modifications to the model's initial error code outputs, thereby generating the correct code. Based on the observations, \selfrefine (FT) is recommended as the more preferable approach for automatic code revision within \ourapproach.

\vspace{6pt}

\RS{5}{The effectiveness of automatic code revision improves with the use of more powerful LLMs. However, using the LLM consistently with the final optimized LLM and under the fine-tuning setting is most beneficial for the final performance.}

\subsection{RQ6: Ablation Study of  \ourapproachbf} \label{RQ6}

\noindent \textbf{Baselines.}
We further perform the ablation study to investigate the effectiveness of each input component in \ourapproach. Requirements and error codes are the indispensable basic inputs for performing automatic code revision. Therefore, we perform ablation experiments on the remaining three components, i.e., \textbf{correct solution}, \textbf{failed test cases}, and \textbf{error messages}. By removing these components individually, we observe their specific impact on the performance of automatic code revision and the final model, and thus evaluate the effectiveness of these components.

\begin{table}[h!]

\caption{Effectiveness of each input component in \ourapproach, where $\mathcal{M}_{\operatorname{Revise}}$ represents the \selfrefine model that used during the training and $\mathcal{M}_{\theta^*}$ represents the final model that used during the inference.}
\centering
{
\begin{tabular}{lccccc}
\toprule
 \multirow{2}{*}{Method} & \multicolumn{3}{c}{\cellcolor{gray!10} $\mathcal{M}_{\operatorname{Revise}}$} & \multicolumn{2}{c}{\cellcolor{gray!20}  $\mathcal{M}_{\theta^*}$} \\ \cmidrule(r){2-4} \cmidrule(r){5-6}
 & Pass@1 & Pass@10 & Pass@any & Pass@1 & Pass@10\\ \midrule
\ourapproach & 3.9\% & 18.9\% & 24.6\% & \textbf{32.8\%} & \textbf{64.0\%}\\
- Correct Solution & 3.4\% & 15.4\% & 19.8\% & 30.1\% & 61.9\%\\
- Error Messages & 3.1\% & 14.2\% & 17.3\% & 28.6\% & 58.7\% \\
- Failed Test Cases & 2.3\% & 5.1\% & 6.3\% & 26.1\% & 47.6\% \\
\bottomrule
\end{tabular}
}\label{ablation_input}
\end{table}

\noindent \textbf{Results.}
We conduct the ablation study on MBPP dataset as shown in Table \ref{ablation_input}. First, we find that removing the failed test cases resulted in the largest drop in performance of all metrics. Failed test cases can demonstrate the inconsistency between the model-generated code output and the desired output, allowing LLMs to reason about and correct erroneous operations. Experimental results show that this point is most helpful for automatic code revision. Second, removing error messages also results in a loss of performance. Error messages directly indicate surface errors in the generated code (such as syntax errors and runtime errors) and the location of the errors, which is also helpful for LLMs to revise the code. The correct code samples in the dataset can provide some reference for revising errors of LLMs, thus reducing the difficulty of correction. We find that the performance of our method without the correct solution drops a little (less than 4\%) compared to the original method. It is still more effective than five existing methods (i.e., Fine-tuning (Full), Fine-tuning (LoRA), Few-shot Prompting, Self-Refine, Self-Debug) as detailed in Table \ref{Baselines}.

\vspace{6pt}

\RS{6}{
The analysis of the ablation study indicates that all of the input components in \selfrefine positively contribute to the effectiveness of automatic code revision and the final model.
}
\section{Related Work}

\subsection{Adaptation of LLMs}
Many tasks rely on adapting LLMs to multiple downstream applications. Such adaptation is usually done via fine-tuning, which updates all the parameters of the pre-trained model. Since LLMs contain a large number of model parameters and performing full parameter tuning would be very expensive, a number of parameter-efficient fine-tuning approaches \cite{prefix_tuning, prompt-tuning, LORA} have been developed.
Adapter tuning \cite{adapter,adapter2} inserts small trainable modules, known as adapters, into pre-trained models. These adapters are usually simple neural networks (e.g., feedforward neural networks) inserted between layers of LLMs. Prompt tuning \cite{prompt-tuning,prompt_tuning2} creates and tunes a small set of artificial tokens, known as a "soft prompt," that is prepended to the input text. These soft prompts are not human-readable text but are instead learned parameters that are optimized during training. Prefix tuning \cite{prefix_tuning,prefix_tuning2} adds a sequence of continuous vectors, known as a "prefix," to the input of each layer of the transformer model. These prefixes are trainable parameters and are optimized during the tuning process. Actually, prefix tuning can also be regarded as deep prompting tuning. Low-rank adaptation \cite{LORA} modifies the LLMs by adding low-rank matrices to the model's parameters. 
These lightweight tuning approaches typically achieve near-full fine-tuning performance while reducing the number of parameters required for training. They primarily optimize the efficiency of training the model parameters but are not directly targeted at improving the efficiency of data sample usage. 

Another approach of adaptation that does not require training is prompting \cite{prompt}, which depends on in-context learning \cite{in-context, in-context2}. By utilizing natural language instructions and a few examples of a task, this approach enables LLMs to recognize and execute new tasks without explicit gradient updates. However, a limitation of this approach is that the models often merely mimic the surface form of the prompt, struggling to deeply understand or adapt to complex and abstract task requirements. Moreover, this approach can be highly sensitive to the specific phrasing of the prompt, leading to variability in the quality and consistency of the generated outputs \cite{PACE}.  

Recent advancements in instruction tuning have leveraged LLMs to synthesize extensive data from initial data, enhancing the ability of LLMs to follow instructions. Works such as self-instruction \cite{Self_Instruct}, code evol-instruct \cite{wizardcoder}, and oss-instruct \cite{Magicoder} approach the problem from the perspective of data augmentation. In contrast, our work focuses on improving the learning efficiency of the models.

Our approach is orthogonal to the aforementioned adaptation techniques, allowing for its concurrent application with these approaches to enhance overall effectiveness.

\subsection{Code Generation with LLM}
The rise of pre-training techniques has brought new momentum to the field of code generation. Against this backdrop, LLMs, such as Codex \cite{codex}, CodeGen \cite{nijkamp2022codegen}, AlphaCode \cite{alphacode}, CodeGeeX \cite{codegeex}, and CodeLlama \cite{CodeLlama} have emerged, greatly enhancing the performance of code generation.  

For LLM-based code generation, there are some methods to refine the outputs produced by LLMs.
Self-refine \cite{Self-Refine} enables LLMs to provide feedback on and correct their own generated content. Self-debug \cite{debug} allows the LLMs to explain and refine their generated code based on execution results. They belong to prompting methods that are constrained by input length and highly sensitive to prompts \cite{sensitive}.
Moreover, Self-edit \cite{Self-Edit} involves training an additional editor.
This category of methods treats refinement as a post-processing step after code generation, whereas we utilize a self-revise to assist the model in efficient training and thereby enhance the model itself. Compared to these post-processing methods, \ourapproach only requires test cases during training. When training is complete, \ourapproach can be directly used without incurring any additional resource or time costs.

Recently, Chen et al. \cite{human_feedback}  propose an ILF method focused on using human feedback to refine model results. 
However, it necessitates continuous human involvement and the provision of feedback throughout the model's training phase, which incurs significant costs in practical applications. 
Chen et al. \cite{Distillation} propose a distillation method that employs ChatGPT \cite{ChatGPT} to generate a large amount of refinement to train small models. However, this method presents two primary limitations. Firstly, it necessitates a highly performant ``teacher'' model, significantly surpassing the capabilities of the ``student'' model. Secondly, commercial constraints and other factors likely prohibit its implementation. 
Furthermore, in parallel to our work, Ding et al. \cite{CYCLE}  propose CYCLE, which enhances the model's self-refinement capability to correct erroneous predictions in previously generated programs according to the test results, rather than focusing on improving the accuracy of a single, one-time prediction, as is the focus of our approach.

In contrast, our work is an adaptation approach, focusing on obtaining a better model adapted to specific domains with limited data. By exploiting the inherent potential of LLMs to achieve error-driven learning, we improve the learning efficiency of LLMs.
\section{Threats to Validity}

There are three major threats to the validity of our work. 

\textbf{1) Threats to external validity} concern the quality of experimental datasets and the generalizability of our results. First, we simulate specific code generation scenarios with five public code generation datasets, which are mainstream benchmarks and have been used in many related works \cite{dataset_previous_1,dataset_previous_2,dataset_previous_3,dataset_previous_4,ROCODE}. Second, \ourapproach can be applied to any LLMs, and we choose four well-known LLMs \cite{model_previous_1,model_previous_2,model_previous_3,model_previous_4} of different sizes, training data, and architectures for our experiments. 

\textbf{2) Threats to internal validity} 
involve the impact of hyperparameters. Deep learning models are known to be sensitive to hyperparameters.For our proposed \ourapproach, we only do a small-range grid search on hyperparameters, including iterations of \ourapproach, learning rates, and training epochs, leaving other hyperparameters the same as those in previous studies \cite{LORA, gpt-3, codegen, llama, llama2, CodeLlama}, which have explored effective settings of the hyperparameters through extensive experiments. For the baselines, their settings are consistent with our approach.

\textbf{3) Threats to construct validity} pertain to the reliability of evaluation metrics. To address this threat, we employ Pass@k \cite{alphacode} as the evaluation metric, which leverages test cases to gauge the functional correctness of code. Additionally, we employ the unbiased version of Pass@k \cite{codex} to diminish evaluation errors that arise from sampling. Pass@k is the mainstream metric for code generation and is widely used in previous studies \cite{CodeScore, pass_at_k_previous_2,pass_at_k_previous_4,Code_Repetition}. On this basis, each experiment is run five times, and its average result is reported.

\section{Limitations}
In this section, we analyze two limitations of \ourapproach as follows.

First, our approach requires test cases. The key point to emphasize is that we only need test cases during the preprocessing stage of training data production. At this stage, we have both the requirements and the ground truth code, which is an important premise. Therefore, during the preprocessing stage, obtaining some test cases is not difficult. We can generate a large number of candidate test cases using traditional methods \cite{EvoSuite} or LLMs \cite{codet} and filter out the correct test cases using the ground truth code \cite{model_previous_3}, which we leave as future work. Compared to test cases, what is truly scarce are the pairs of requirements and ground truth code. Our approach focuses on improving the model's performance in this more urgent need. 

Second, \ourapproach is only used in low-resource scenarios. Limited training data is a realistic and key problem existing in the real world, which is usually more difficult to solve. In this paper, we try to give an effective approach to address this problem and demonstrate that our approach can achieve significant improvements in the experiments of low-resource settings.

\section{Conclusion}
In this work, we have proposed \ourapproach, a Data-Efficient adaptation with Error-Driven learning for code generation, improving the code generation performance of LLMs in specific scenarios with limited samples. \ourapproach outperforms mainstream adaptation approaches (e.g., full-parameter fine-tuning, LoRA, and few-shot prompting) with limited training data on five code generation benchmarks. Our analysis indicates that LLMs can learn more efficiently by learning from the revisions of their error than by traditionally learning from code examples in datasets.
Extensive results show that \ourapproach can largely enhance five different LLMs of varying sizes or series, which proves its applicability. We believe that improving the LLMs' learning efficiency and adapting LLMs with fewer samples has a broad application in real-world scenarios. We leave this area of exploration for future work.

\begin{acks}
This research is supported by the National Key R\&D Program under Grant No. 2023YFB4503801, 
the National Natural Science Foundation of China under Grant No. 62192733, 62192730, 62192731, and the Major Program (JD) of Hubei Province (No.2023BAA024)
\end{acks}

\newpage
\appendix

\section{Effect of Training Data Size on Performance}

\begin{figure*}[h!]
\centering
\color{blue}
\includegraphics[width=0.58\textwidth]{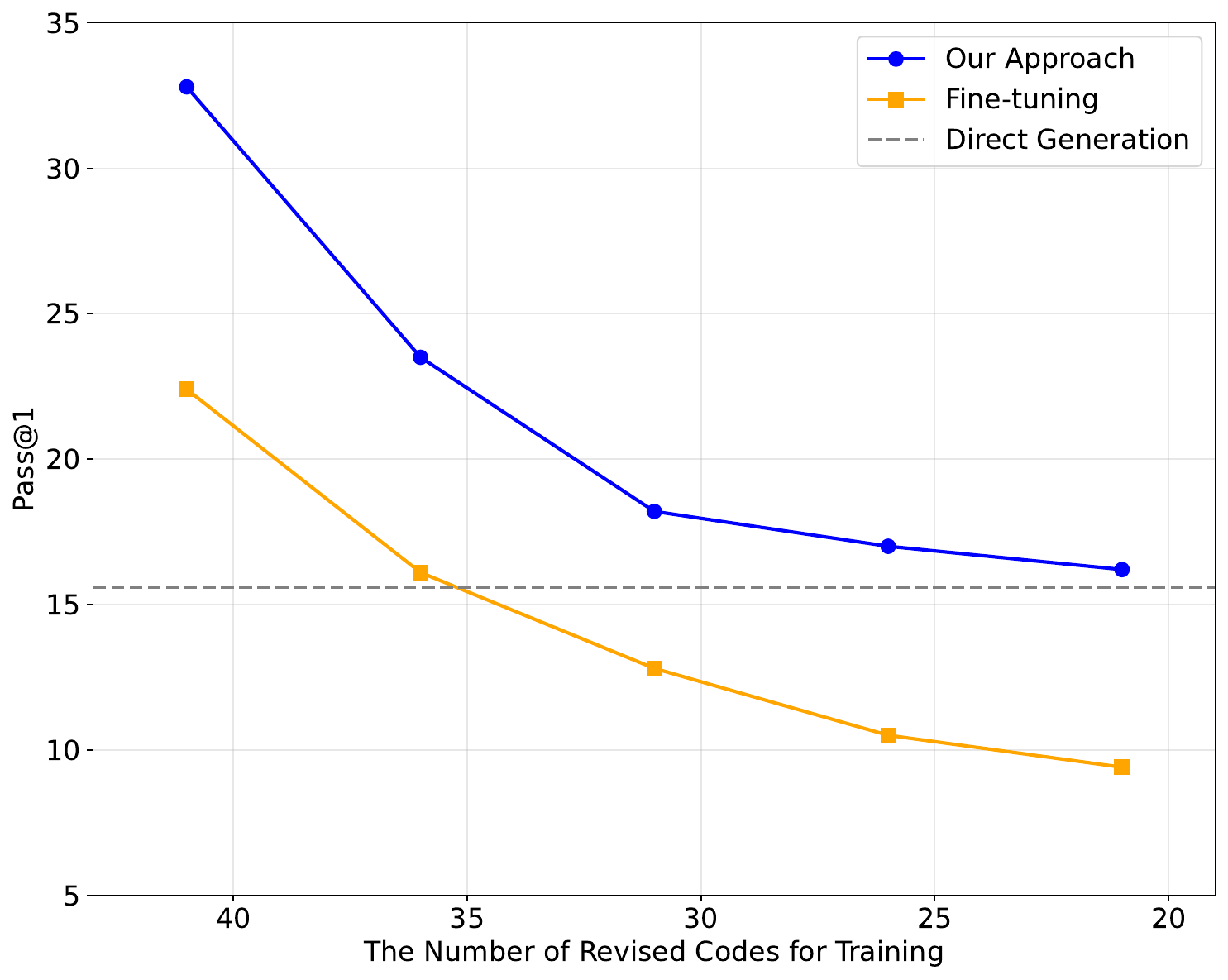}
\centering
\caption{Performance analysis with varying sizes of training data on MBPP dataset.}
\label{vary_revised}
\end{figure*}

We investigate the effect of training data size on performance. This experiment is conducted on MBPP dataset, and the results are presented in Figure \ref{vary_revised}.
As the amount of training data decreases, our approach consistently outperforms both the Direct Generation with base model and the Fine-tuning method. The Fine-tuning method exhibits a much steeper performance decline as data is reduced, quickly underperforming the Direct Generation, which suggests it is more prone to overfitting. However, our method can learn effectively from scarce data.

Furthermore, we also find a useful case study that helps illustrate the learning capabilities of our approach. One error pattern in the MBPP data is that the model (CodeGen-2B) fails to import Python libraries used in the code. We observe that even after removing all training samples that corrected for a specific missing library, the model can still learn to import it by generalizing from samples involving other libraries. This suggests that the model can learn more generalizable error patterns from existing data.

\section{Project-level code generation with domain-specific evaluations}

We conduct a further evaluation on Evocodebench \cite{Evocodebench}. Evocodebench is a project-level code generation dataset designed to address the problem of data leakage \cite{CDD} in evaluation. It achieves this by continuously evolving its data from real-world software projects, while also focusing on domain-specific assessments.
Following the original benchmark's methodology, we use the natural language (NL) requirements, function signatures, and surrounding code as prompts. Performance is then measured against the provided test cases using the Pass@k metric. For this evaluation, we use the "camp zifnerf" project of Evocodebench, which contains 49 scientific and engineering samples, and randomly split them into training and test sets at a 2:1 ratio.

\begin{table}[h]
\centering
\begin{tabular}{lccc}
\toprule
Method & \textbf{Pass@1} & \textbf{Pass@5} & \textbf{Pass@10} \\
\midrule
Direct Generation & 24.0\% & 46.7\% & 50.0\% \\
Fine-tuning & 26.7\% & 44.4\% & 46.7\% \\
\midrule
\textbf{DEED} & \textbf{28.0\%} & \textbf{51.1\%} & \textbf{53.3\%} \\
\bottomrule
\end{tabular}
\caption{Performance of \ourapproach on Evocodebench dataset.}
\label{tab:results_Evocodebench2}
\end{table}

The results are shown in Table \ref{tab:results_Evocodebench2}. Our method outperforms the Direct Generation and Fine-tuning baselines across Pass@1, Pass@5, and Pass@10 metrics. This result demonstrates the effectiveness of our approach in real data-scarce scenarios. Furthermore, this performance indicates \ourapproach's broad applicability, proving its efficacy for domain-specific, project-level tasks.

\section{The Instruction for Automatic Code Revision}
We use a reasonable instruction at the beginning of the prompt for automatic code revision. All baselines use the same prompt. The specific instruction is as follows:

\vspace{0.5cm}

\begin{minipage}{0.95\linewidth}
\texttt{%
"I've encountered an error code. I will show you the correct code snippet and ask for your assistance in fixing the error based on that correct code with minimal necessary revisions."
}
\end{minipage}%

\bibliographystyle{ACM-Reference-Format}
\bibliography{ref}

\end{document}